\documentclass[aps,pre,preprint]{revtex4}
\usepackage{graphicx}
\begin{document}

\title{Three-dimensional instability of two nonlinearly coupled electromagnetic waves in a plasma}
\author{Lennart Stenflo, Bengt Eliasson, and Mattias Marklund}
\affiliation{Department of Physics, Ume{\aa} University, SE--901 87 Ume{\aa}, Sweden}

\received{10 May 2007} \revised{15 May 2007; Accepted 15 May 2007}

\begin{abstract}
The three-dimensional instability of two coupled electromagnetic
waves in an unmagnetized plasma is investigated theoretically and
numerically. In the regime of two-plasmon decay, where one pump wave
frequency is approximately twice the electron plasma frequency, we
find that the coupled pump waves give rise to enhanced instability
with wave vectors between those of the two beams. In the case of ion
parametric decay instability, where the pump wave decays into one
Langmuir wave and one ion acoustic wave, the instability regions are
added with no distinct amplification. Our investigation can be
useful in interpreting laser-plasma as well as ionospheric heating
experiments.
\end{abstract}

\maketitle

\section{Introduction}

The development of high-power lasers has made it possible to study
the relativistic regime of laser-plasma interactions. This is of
fundamental importance for laser-induced heating and inertial
confinement fusion \cite{Kodama01} and poses also the possibility to
test fundamental physics. The first theoretical studies of
relativistic waves in plasmas date back to the fifties
\cite{Akhiezer56}, where it was recognized that the the quivering
velocity of electrons may lead to relativistic mass increase in an
ultra-intense electromagnetic wave. The relativistic effects can
result in self-modulation and self-focusing of electromagnetic waves
in plasmas \cite{Max74}. The instability of relativistically large
amplitude electromagnetic waves has been studied for magnetized
electron plasmas \cite{Stenflo76}, electron-ion plasmas
\cite{Stenflo84,Shukla86}, electron beam plasma systems
\cite{Tsintsadze74}, hot electron-ion plasmas \cite{Tsintsadze79},
and magnetized hot plasmas \cite{Tsintsadze80,Phys_Rep}. The
instability of relativistically strong laser light in an
unmagnetized plasma has then been revisited \cite{Sakharov94}. The
presence of multiple laser beams in a plasma can give rise to a new
set of interesting phenomena \cite{Rosenbluth72,
Bingham04,McKinstrie92,Shukla92}. An important application of two
interacting laser beams in plasmas is the excitation of large
amplitude Langmuir waves, which in turn accelerate electrons to
ultra-relativistic speeds \cite{Bingham04}. The nonlinear coupling
between two electromagnetic waves in plasmas can be described by a
system of coupled nonlinear Schr\"odinger equations that model
nonlinear interactions between localized light
\cite{Shukla92,Berge98} and Langmuir or ion-acoustic waves. At
strongly relativistic intensities, laser beams can give rise to fast
plasma waves via higher-order nonlinearities
\cite{Rosenbluth72,Shvets01,Bingham04}, or via the beat wave
excitation at frequencies different from the electron plasma
frequency \cite{Shvets04}. Particle-in-cell simulations
\cite{Nagashima01} show that large-amplitude electron plasma waves
can be excited by colliding laser pulses, or by two co-propagating
electromagnetic pulses where a long trailing pulse is modulated
efficiently by the periodic plasma wake behind the heading short
laser pulse \cite{Sheng02}. The effects on parametric instabilities
of a partially incoherent pump wave (with a distribution of wave
modes) was investigated both theoretically \cite{Thomson75} and
experimentally \cite{Obenschain76}, where it was found that the
effect of finite bandwidth is, in general, to increase the
instability thresholds and lower the growth rate. The nonlinear
interaction between two electromagnetic waves in the Earth's
ionosphere has also been considered \cite{r12,r13,r14}. In that
case, it is, in addition, important to focus attention on the
collisional coupling between the waves, e.g. \cite{r5,r6,r7}.
Related phenomena occur also in semiconductor plasmas \cite{r8,r8b}.

Our previous treatment of the instability of two coupled laser beams
in an electron-ion plasma \cite{Shukla06} and a two-temperature
electron plasma \cite{Eliasson06} were limited to the investigation
of the relativistic Raman and Brillouin scattering instabilities in
one dimension, and two-dimensional effects were only partly
included. The purpose of the present paper is to include the
multi-dimensional effects, which are particularly important for the
cases where an electromagnetic wave decays into one low-frequency
wave and one high-frequency electrostatic wave that are propagating
obliquely to the pump wave.

\section{Model equations}

Let us consider the propagation of intense laser light in an electron--ion plasma. The
dynamics of the high frequency laser light is governed by
\begin{equation}
  \frac{\partial^2\mathbf{A}_h}{\partial t^2} + c^2\nabla\times(\nabla\times\mathbf{A}_h)
    - 3v_{Te}^2\nabla(\nabla\cdot\mathbf{A}_h) + \omega_{pe}^2(1 + N_{es}) \mathbf{A}_h
    - \frac{\omega_{pe}^2e^2}{m_e^2c^4}\langle|\mathbf{A}_h|^2\rangle\mathbf{A}_h = 0 ,
\end{equation}
where $\omega_{pe} = (4\pi n_0e^2/m_e)^{1/2}$ is the electron plasma frequency,
$n_0$ is the equilibrium number density, $e$ is the magnitude of the electron charge,
$m_e$ is the electron mass, $v_{Te}$ is the electron thermal speed,
$c$ is the speed of light, and the normalized slow time-scale electron number
density perturbation is $N_{es} = n_{es}/n_0$.
The latter is excited by the ponderomotive force of the high frequency waves.
If the ions are considered as immobile, we have \cite{Shukla06}
\begin{equation}
  \left(
    \frac{\partial^2}{\partial t^2} + \omega_{pe}^2 - 3v_{Te}^2\nabla^2
  \right)N_{es} = \frac{e^2}{m_e^2c^2}\nabla^2\langle|\mathbf{A}_h|^2\rangle
\end{equation}
whereas if the electrons are treated as inertialess \cite{Shukla06}
\begin{equation}
  \left(
    \frac{\partial^2}{\partial t^2} - c_{s}^2\nabla^2
  \right)N_{es} = \frac{e^2}{m_em_ic^2}\nabla^2\langle|\mathbf{A}_h|^2\rangle
\end{equation}
where $c_s$ is the sound speed.

We will here consider two large amplitude electromagnetic waves $\mathbf{A}_h = \mathbf{A}_1 + \mathbf{A}_2$.
We then have $|\mathbf{A}_h|^2 = |\mathbf{A}_1|^2 + |\mathbf{A}_2|^2 + 2\mathbf{A}_1\cdot\mathbf{A}_2$.

Using equations (1)--(3) we can now derive a nonlinear dispersion relation. The calculations are
straightforward but lengthy. Following closely the analysis of Ref.\ \cite{r9}, it then turns out
to be convenient to introduce a characteristic velocity $v_{tS}$ that is defined by
\begin{equation}
  v_{tS,j}^2 = \omega_{pe}^2\sum_{+,-}\left[
    \frac{|\mathbf{k}_{j \pm}\times \mathbf{v}_{0j}|^2}{k_{j\pm }^2D_{j \pm}}
    + \frac{|\mathbf{k}_{j \pm}\cdot\mathbf{v}_{0j}|^2}{k_{j\pm }^2\omega_{j \pm}^2
      \varepsilon_{j\pm}}
  \right]
\end{equation}
where $\omega_{j \pm} = \omega \pm \omega_{0j}$, $\mathbf{k}_{j \pm} = \mathbf{k} \pm \mathbf{k}_{0 j}$,
$\omega_{0j}$ and $\mathbf{k}_{0 j}$ is the pump frequency and wavevector of the pump wave $j$
($j = 1,\,2$), whereas $\omega$ and $\mathbf{k}$ are associated with the electrostatic low-frequency
fluctuations. Furthermore, we have here introduced the pump wave quiver velocity
${\bf v}_{0j}=e{\bf A}_{0j}/m_e c$, $D_{j\pm} = \omega_{j\pm}^2
- \omega_{pe}^2  -k_{j \pm}^2c^2 + i\omega_{j\pm}\gamma_{j\pm}$ where $\gamma_{j\pm}$ represents the
sideband damping \cite{r9}, and $ \omega_{j\pm}^2 \varepsilon_{j\pm} \approx \omega_{j\pm}^2 -
\omega_{pe}^2-3 {k}_{j\pm}^2v_{Te}^2$. For the pump frequencies, we use
$\omega_{0j}^2=\omega_{pe}^2+k_{0j}^2c^2$.

In the present paper we will, for simplicity, further
assume that $\mathbf{A}_1\cdot\mathbf{A}_2\approx 0$.
This means that double resonance parametric phenomena \cite{r10},
where the difference
between the frequencies $\omega_{01}$ and $\omega_{02}$ is close to twice a natural frequency, will be neglected.
Choosing two transverse waves that propagate in the $y-$ and $z$-directions, respectively
with the pump velocities in the $z-$ and $y$-directions, respectively, we note that (4) reduces to
\begin{equation}
  v_{tS, j}^2 \approx \frac{\omega_{pe}^2e^2}{m_e^2c^2}\left[
\frac{|\mathbf{k}_{j+}\times\mathbf{A}_{0j}|^2}{{k}_{j+}^2 D_{j+}}
  + \frac{|\mathbf{k}_{j-}\times\mathbf{A}_{0j}|^2}{{k}_{j-}^2 D_{j-}}
  + \frac{|\mathbf{k}\cdot\mathbf{A}_{0j}|^2}{{k}_{j+}^2\omega_{j+}^2\varepsilon_{j+}}
  + \frac{|\mathbf{k}\cdot\mathbf{A}_{0j}|^2}{{k}_{j-}^2\omega_{j-}^2\varepsilon_{j-}}
  \right]
\end{equation}
With these limitations, and following Refs. \cite{Shukla06} and \cite{r9}, the nonlinear dispersion relation turns out to be
\begin{eqnarray}
  &&
  \frac{1}{Q} +
    \frac{|\mathbf{k}_{1+}\times\mathbf{A}_{01}|^2}{{k}_{1+}^2 D_{1+}}
  + \frac{|\mathbf{k}_{1-}\times\mathbf{A}_{01}|^2}{{k}_{1-}^2 D_{1-}}
  + \frac{|\mathbf{k}_{2+}\times\mathbf{A}_{02}|^2}{{k}_{2+}^2 D_{2+}}
  + \frac{|\mathbf{k}_{2-}\times\mathbf{A}_{02}|^2}{{k}_{2-}^2 D_{2-}}
  \nonumber \\[2mm] &&\quad
  + \frac{|\mathbf{k}\cdot\mathbf{A}_{01}|^2}{{k}_{1+}^2\omega_{1+}^2\varepsilon_{1+}}
  + \frac{|\mathbf{k}\cdot\mathbf{A}_{01}|^2}{{k}_{1-}^2\omega_{1-}^2\varepsilon_{1-}}
  + \frac{|\mathbf{k}\cdot\mathbf{A}_{02}|^2}{{k}_{2+}^2\omega_{2+}^2\varepsilon_{2+}}
  + \frac{|\mathbf{k}\cdot\mathbf{A}_{02}|^2}{{k}_{2-}^2\omega_{2-}^2\varepsilon_{2-}}
  = 0  .
\label{dispersion}
\end{eqnarray}
In Eq. (6), as well as below, we have normalized ${\bf A}_{0j}$ by
$m_e c^2/e$.
For electron Langmuir waves, we have \cite{Shukla06}
\begin{equation}
Q_L =\omega_{pe}^2\left(
1-\frac{k^2c^2}{\omega^2-3k^2v_{Te}^2-\omega_{pe}^2}
\right).
\label{Langmuir}
\end{equation}
whereas for ion acoustic waves \cite{Shukla06}
\begin{equation}
  Q_{IA} = \omega_{pe}^2\left(
 1-\frac{m_e}{m_i}\,\frac{k^2 c^2}{(\omega^2-k^2 c_s^2)}
\right).
\label{ion}
\end{equation}

\section{Results}

\begin{figure}
\centering
\includegraphics[width=10cm]{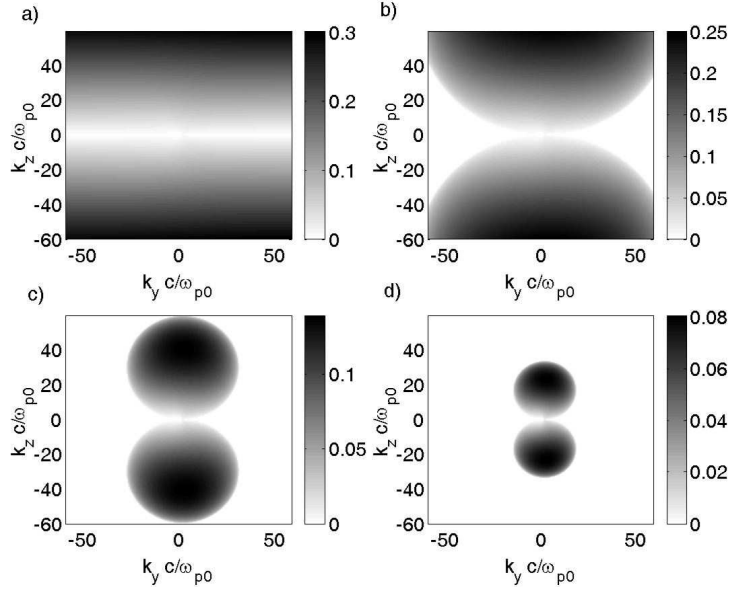}
\caption{The growth rate (normalized by $\omega_{pe}$) of the two-plasmon decay
of one laser beam for different electron thermal
speeds a) $v_{Te}=0$, b) $v_{Te}=0.005\,\mathrm{c}$, c) $v_{Te}=0.0075\,\mathrm{c}$,
and d) $v_{Te}=0.01\,\mathrm{c}$. The pump amplitude is ${\bf A}_{01}=0.01\,\widehat{\bf z}$, the pump wavevector
$\mathbf{k}_{01}=\sqrt{3}\widehat{\bf y}\omega_{pe}/\mathrm{c}$.
The second laser beam intensity is set to zero (${\bf A}_{02}={\bf 0}$).
}
\end{figure}

\begin{figure}
\centering
\includegraphics[width=10cm]{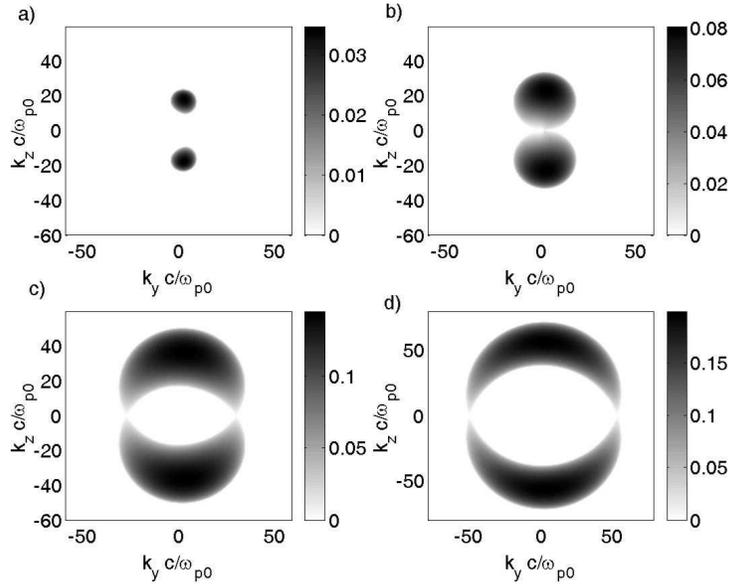}
\caption{The growth rate (normalized by $\omega_{pe}$) of the two-plasmon decay of one laser beam for wavevectors
a) $\mathbf{k}_{01}=1.65\widehat{\bf y}\omega_{pe}/\mathrm{c}$,
b) $\mathbf{k}_{01}=\sqrt{3}\widehat{\bf y}\omega_{pe}/\mathrm{c}$,
c) $\mathbf{k}_{01}=2\widehat{\bf y}\omega_{pe}/\mathrm{c}$,
and d) $\mathbf{k}_{01}=2.5\widehat{\bf y}\omega_{pe}/\mathrm{c}$.
The pump amplitude is ${\bf A}_{01}=0.01\,\widehat{\bf z}$, and the electron thermal speed
$v_{Te}=0.01\,\mathrm{c}$. The second laser beam intensity is set to zero (${\bf A}_{02}={\bf 0}$).
}
\end{figure}

\begin{figure}
\centering
\includegraphics[width=10cm]{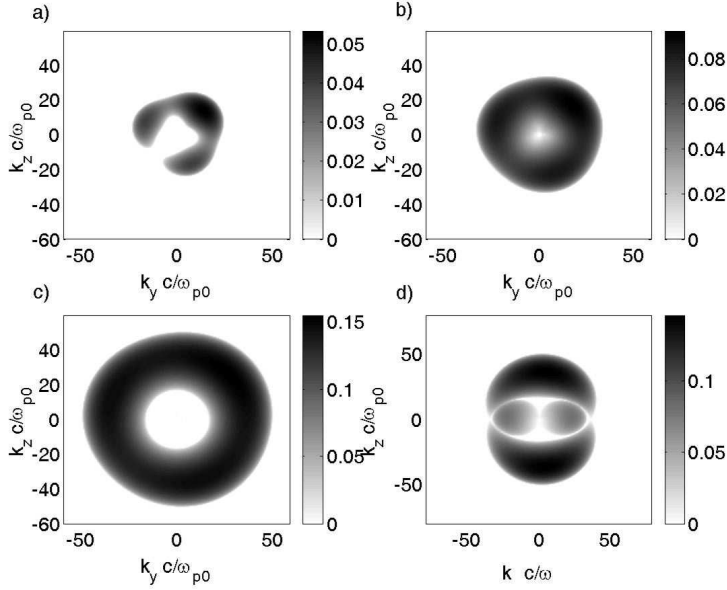}
\caption{The growth rate (normalized by $\omega_{pe}$) of the two-plasmon decay
of two coupled laser beams for wavevectors
a) $\mathbf{k}_{01}=1.65\widehat{\bf y}\omega_{pe}/\mathrm{c}$ and $\mathbf{k}_{02}=1.65\widehat{\bf z}\omega_{pe}/\mathrm{c}$,
b) $\mathbf{k}_{01}=\sqrt{3}\widehat{\bf y}\omega_{pe}/\mathrm{c}$ and $\mathbf{k}_{02}=\sqrt{3}\widehat{\bf z}\omega_{pe}/\mathrm{c}$,
c) $\mathbf{k}_{01}=2\widehat{\bf y}\omega_{pe}/\mathrm{c}$ and $\mathbf{k}_{02}=2\widehat{\bf z}\omega_{pe}/\mathrm{c}$,
and d) $\mathbf{k}_{01}=2\widehat{\bf y}\omega_{pe}/\mathrm{c}$ and $\mathbf{k}_{02}=\sqrt{3}\widehat{\bf z}\omega_{pe}/\mathrm{c}$.
The pump amplitudes are ${\bf A}_{01}=0.01\,\widehat{\bf z}$
and ${\bf A}_{02}=0.01\,\widehat{\bf y}$,
and the electron thermal speed is
$v_{Te}=0.01\,\mathrm{c}$.
}
\end{figure}

\begin{figure}
\centering
\includegraphics[width=10cm]{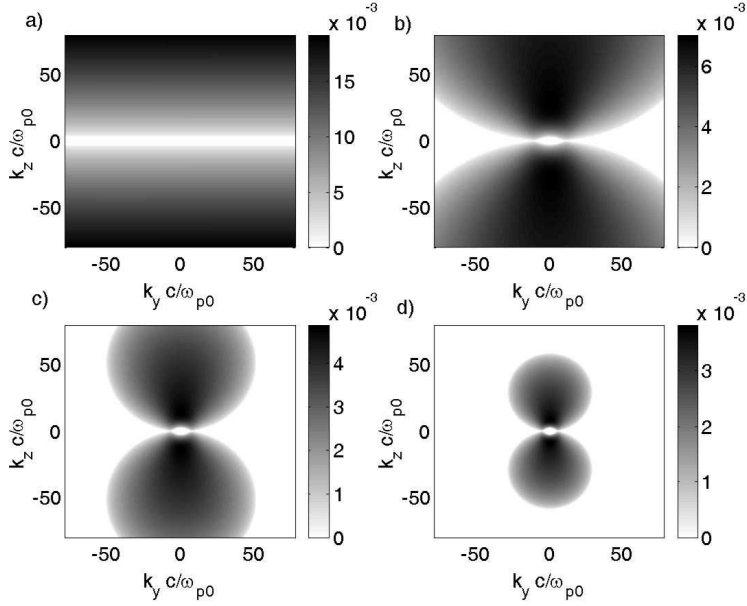}
\caption{The growth rate (normalized by $\omega_{pe}$) of the ion parametric decay instability
of one laser beam for different electron thermal
speeds a) $v_{Te}=0$, b) $v_{Te}=0.005\,\mathrm{c}$, c) $v_{Te}=0.0075\,\mathrm{c}$,
and d) $v_{Te}=0.01\,\mathrm{c}$; in each case the ion acoustic speed is set to $c_s=0.006\,v_{Te}$.
The pump amplitude is ${\bf A}_{01}=0.01\,\widehat{\bf z}$, and the pump wavevector is
$\mathbf{k}_{01}=0.1\widehat{\bf y}\omega_{pe}/\mathrm{c}$.
The second laser beam intensity is set to zero (${\bf A}_{02}={\bf 0}$).
}
\end{figure}

\begin{figure}
\centering
\includegraphics[width=10cm]{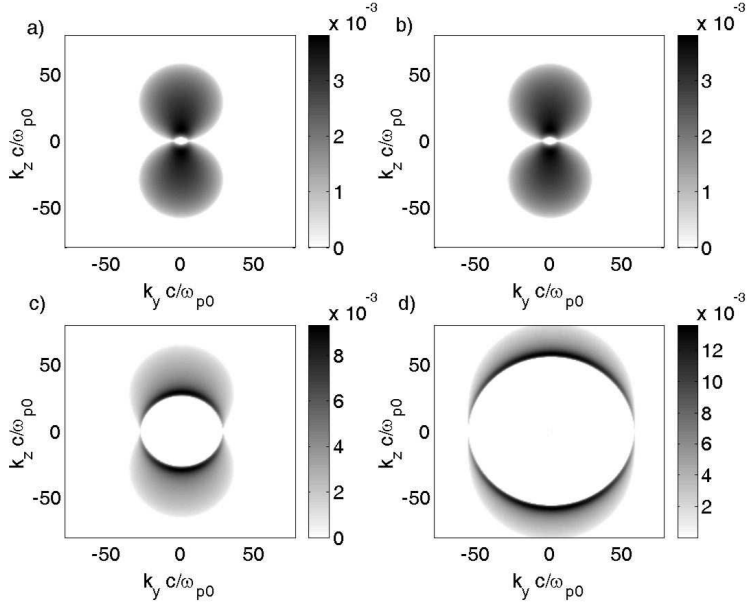}
\caption{The growth rate (normalized by $\omega_{pe}$) of the ion parametric decay instability of one laser beam for wavevectors
a) $\mathbf{k}_{01}={\bf 0}$,
b) $\mathbf{k}_{01}=0.1\widehat{\bf y}\omega_{pe}/\mathrm{c}$,
c) $\mathbf{k}_{01}=0.5\widehat{\bf y}\omega_{pe}/\mathrm{c}$,
and d) $\mathbf{k}_{01}=\widehat{\bf y}\omega_{pe}/\mathrm{c}$.
The pump amplitude is ${\bf A}_{01}=0.01\,\widehat{\bf z}$, the electron thermal speed is
$v_{Te}=0.01\,\mathrm{c}$, and the ion acoustic speed is $c_s=6\times10^{-5} c$.
The second laser beam intensity is set to zero (${\bf A}_{02}={\bf 0}$).
}
\end{figure}

\begin{figure}
\centering
\includegraphics[width=10cm]{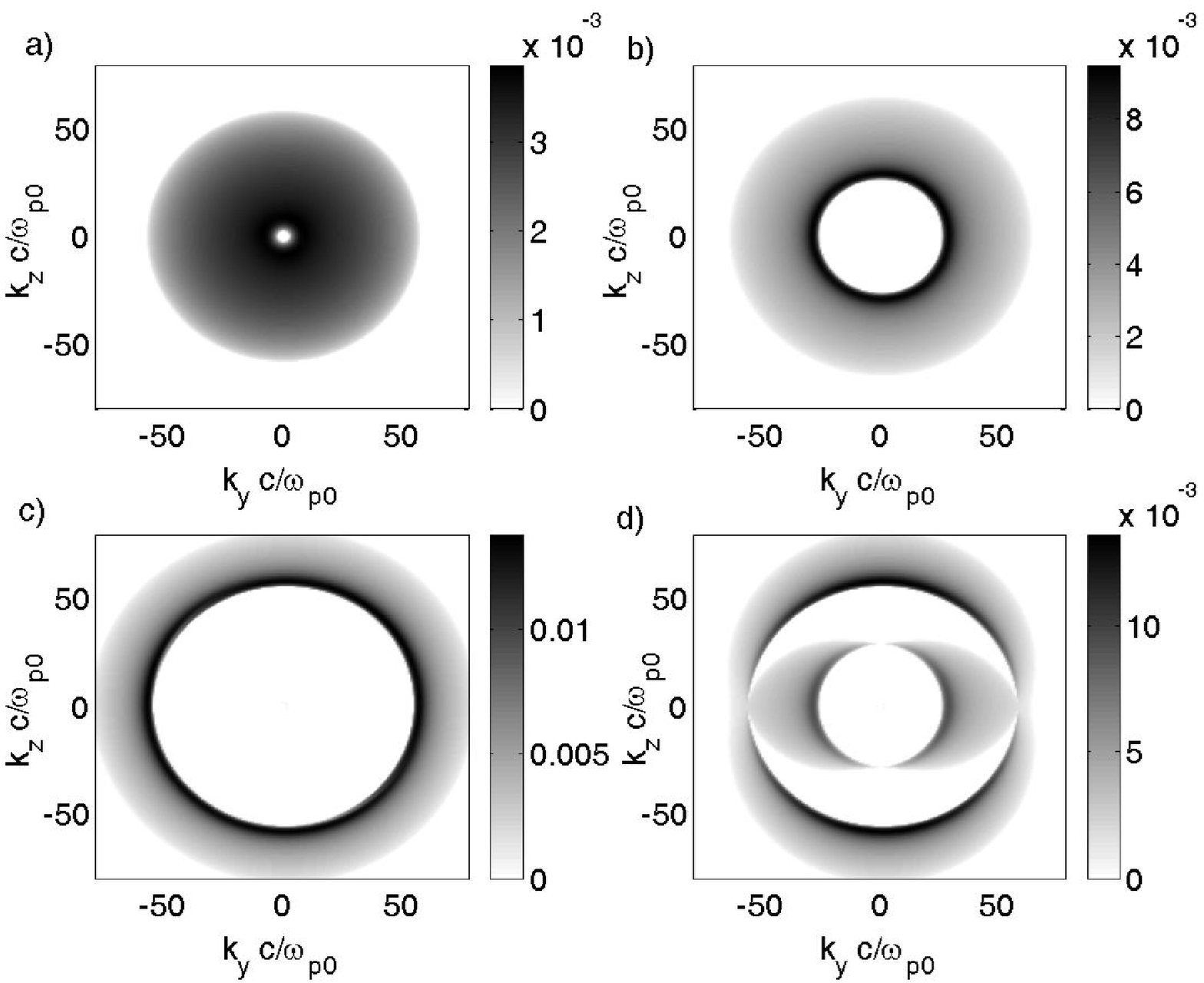}
\caption{The growth rate (normalized by $\omega_{pe}$) of the ion parametric decay instability
of two coupled laser beams for wavevectors
a) $\mathbf{k}_{01}=0.1\widehat{\bf y}\omega_{pe}/\mathrm{c}$ and $\mathbf{k}_{02}=0.1\widehat{\bf z}\omega_{pe}/\mathrm{c}$,
b) $\mathbf{k}_{01}=0.5\widehat{\bf y}\omega_{pe}/\mathrm{c}$ and $\mathbf{k}_{02}=0.5\widehat{\bf z}\omega_{pe}/\mathrm{c}$,
c) $\mathbf{k}_{01}=\widehat{\bf y}\omega_{pe}/\mathrm{c}$ and $\mathbf{k}_{02}=\widehat{\bf z}\omega_{pe}/\mathrm{c}$,
and d) $\mathbf{k}_{01}=\widehat{\bf y}\omega_{pe}/\mathrm{c}$ and $\mathbf{k}_{02}=0.5\widehat{\bf z}\omega_{pe}/\mathrm{c}$.
The pump amplitudes are ${\bf A}_{01}=0.01\,\widehat{\bf z}$
and ${\bf A}_{02}=0.01\,\widehat{\bf y}$,
the electron thermal speed is $v_{Te}=0.01\,\mathrm{c}$ and the ion acoustic speed is $c_s=6\times10^{-5} c$.
}
\end{figure}

We have carried out a numerical analysis of Eq. (\ref{dispersion}), where we have assumed that the
frequency is complex valued, where the imaginary part of $\omega$ represents the growth rate.
In our treatment, we have concentrated on three-wave decay processes where the important
terms in Eq.~(\ref{dispersion}) are the ones with down-shifted daughter waves. Hence, we have
kept the resonant terms with subscripts "-" in Eq. (\ref{dispersion}) but neglected those with subscripts
"+". For large wavenumbers, we recover almost identically
the stimulated Raman and Brillouin cases treated in Ref. \cite{Shukla06}. Thus, we shall here concentrate
on the regimes of smaller wavenumbers in which the two-plasmon decay and ion parametric
decay instabilities become important, and where it is crucial to have a fully
multi-dimensional treatment.

We first consider the two-plasmon case in which the low-frequency wave is a Langmuir wave, and where $Q$
in Eq. (\ref{dispersion}) equals $Q_L$ in (\ref{Langmuir}).
The growth rates (normalized by $\omega_{pe}$) are presented in Figs. 1--3.
We have here denoted the unit vectors in the $y$ and $z$ directions by
$\widehat{\bf y}$ and $\widehat{\bf z}$. In Fig. 1,
the electron thermal effects are investigated. The dispersion of the Langmuir waves due
to the electron pressure causes the instability to restricted in a bounded
domain in wavevector space, with a maximum growth rate in a direction perpendicular
to the propagation direction of the laser beam. In Fig. 2, we have investigated how the
instability depends on the pump wavenumber of a single electromagnetic beam. We find that
the instability is dominant for wavenumbers equal to or larger than $\sqrt{3}\omega_{pe}/c$,
and that the instability vanishes for smaller wavenumbers of the pump wave.
The maximum instability occurs for wavevectors almost perpendicular to the laser beam propagation direction.
The instability maximum occurs for wavenumbers much larger than the pump wavenumber.
Eventually kinetic effects will become
important and electron Landau damping will decrease the growth rate. The case of two
coupled electromagnetic beams is investigated in Fig. 3. While beam 1 is directed
in the $y$-direction as in Figs. 1 and 2, beam $2$ is here in the $z$-direction,
perpendicular to beam 1. We see in panel a) of Fig. 3 that the coupled beam
system gives rise to a new instability with a maximum growth rate in the direction
of the dichotome in the center between the two beam propagation directions.
For wavenumbers larger than $\sqrt{3}\omega_{pe}/c$, the instability becomes
more evenly distributed in all directions, but with a well-defined maximum
growth rate for some wavenumber. For the
case where the wavenumber of beam 2 is smaller than the one of beam 1, in panel d), we see
a superposition of the two instability regions for the separate beams.
We note that similar effects can also appear due to direct subharmonic wave generation in
nonuniform plasmas \cite{r15}.

We next investigate the parametric decay instability in which the electromagnetic
wave decays into one electrostatic wave and one low-frequency ion
acoustic wave, where $Q$ in Eq. (\ref{dispersion}) equals $Q_{IA}$ in (\ref{ion}).
The growth rates are presented in Figs. (4)--(6). We have here concentrated on the long
wavelength limit where the main instability is the ion parametric decay instability,
in which the electromagnetic wave decays into one slightly frequency-downshifted
electrostatic wave and one low-frequency ion acoustic wave. In Fig. 4, we have studied
the thermal effect on the instability of one single electromagnetic beam. We see
that for higher electron thermal and ion acoustic speeds, the region of instability
becomes smaller in wavenumber space, and that there is a well-defined maximum
of the instability for propagation almost perpendicular to the wave propagation direction. In
Fig. 5, we have considered different wavenumbers of the pump wave. We see that for
large wavenumbers, the maximum instability occurs for larger wavevectors perpendicular
to the pump wavevector. Finally, we consider the interaction between two coupled
electromagnetic beams in Fig. 6. For the cases of equal pump amplitudes and
lengths of the wavevectors, in panels a)--c), the instability becomes almost
rotationally symmetric, with equal maximum growth rates in all directions. For the
case where the wavenumber of beam 2 is smaller than the one of beam 1, in panel d), we see
a superposition of the two instability regions for the separate beams. For the ion
parametric decay instability, we could not see the same distinct amplification of
the instabilities as we could see for the two-plasmon decay in panel a) of Fig. 3.

In conclusion, we have investigated the instability of two coupled large-amplitude electromagnetic
waves in a plasma. Our investigation shows that two-plasmon decay plays an important
role for pump frequencies approximately two times larger than the background electron plasma frequency,
and that one has a maximum growth rate perpendicular to the propagation direction of the electromagnetic
wave. For two coupled electromagnetic beams, there is here a new and stronger instability in a direction
between the two laser beams. The ion parametric decay instability, in which an electromagnetic
wave decays into one Langmuir wave and one ion acoustic wave, is important
for long wavelengths of the electromagnetic pump wave, where the instability leads
to ion acoustic waves that are propagating perpendicularly to the electromagnetic beam
direction. Here, the addition of a second electromagnetic beam leads to a superposition of the
instabilities. However, we see no strong nonlinear amplification
due to the two beams.  Our study could be important for both laser-plasma interactions
and for ionospheric heating experiments \cite{r16a,r16,r17,r18}.

\end{document}